How relevant is climate change research for climate change policy?

An empirical analysis based on Overton data

Lutz Bornmann*$, Robin Haunschild$, Kevin Boyack**, Werner Marx$, Jan C. Minx***

\* Science Policy and Strategy Department

Administrative Headquarters of the Max Planck Society

Hofgartenstr. 8,

80539 Munich, Germany.

Email: bornmann@gv.mpg.de

$ Max Planck Institute for Solid State Research

Heisenbergstraße 1,

70569 Stuttgart, Germany.

Email: r.haunschild@fkf.mpg.de, w.marx@fkf.mpg.de, l.bornmann@fkf.mpg.de

\*\* SciTech Strategies, Inc.

8421 Manuel Cia Pl NE

Albuquerque, NM 87122, USA.

Email: kboyack@mapofscience.com

\*\*\* Mercator Research Institute on Global Commons and Climate Change (MCC)

Torgauer Str. 12-15,

10829 Berlin, Germany.

Email: minx@mcc-berlin.net

**Abstract**

Climate change is an ongoing topic in nearly all areas of society since many years. A discussion of climate change without referring to scientific results is not imaginable. This is especially the case for policies since action on the macro scale is required to avoid costly consequences for society. In this study, we deal with the question of how research on climate change and policy are connected. In 2019, the new Overton database of policy documents was released including links to research papers that are cited by policy documents. The use of results and recommendations from research on climate change might be reflected in citations of scientific papers in policy documents. Although we suspect a lot of uncertainty related to the coverage of policy documents in Overton, there seems to be an impact of international climate policy cycles on policy document publication. We observe local peaks in climate policy documents around major decisions in international climate diplomacy. Our results point out that IGOs and think tanks – with a focus on climate change – have published more climate change policy documents than expected. We found that climate change papers that are cited in climate change policy documents received significantly more citations on average than climate change papers that are not cited in these documents. Both areas of society (science and policy) focus on similar climate change research fields: biology, earth sciences, engineering, and disease sciences. Based on these and other empirical results in this study, we propose a simple model of policy impact considering a chain of different document types: the chain starts with scientific assessment reports (systematic reviews) that lead via science communication documents (policy briefs, policy reports or plain language summaries) and government reports to legislative documents.

# 1 Introduction

People have long believed that nature is so vast and powerful that mankind has not the potential for any major and lasting effect on the earth's climatic system. One century ago, Arrhenius (1896), one of the discoverers of the greenhouse effect, even welcomed a hotter climate for Northern Europe. According to Weart (2008), the World Climate Conference in Geneva in 1979 and the reports of the US National Academy of Sciences (NAS) and the US Environmental Protection Agency (EPA) in 1983 are important milestones at the beginning of the climate debate, particularly beyond the scientific community.

In the 1960s many experts assumed that swings of the global mean temperature take tens of thousands of years; in the 1970s, they assumed thousands of years. Meanwhile, ice core data from the last Glacial Period show that abrupt global warming is possible and can happen within a few decades or even within a few years as a climate shock (see IPCC, 2014, p. 16, climate change beyond 2100, irreversibility and abrupt changes). In the 1980s, climate change was no longer a theoretical problem. It was widely agreed among experts that global warming could be a concrete threat. A growing number of well-respected climate researchers (like Roger Revelle, Stephen Schneider, James Hansen, Bert Bolin) were deeply concerned and pointed out that the earth was getting noticeably warmer. A series of meetings of meteorologists held in Villach, Austria, led to a growing conviction that global warming may not be a problem of the far future but might become serious within the scientists' own lifetimes. Subsequently, scientists took an active stance and prompted governments to act soon, because the rate and degree of future warming could be influenced by governmental policy (see Weart, 2008, pp. 138-154, breaking into policy).

The year 1988 marked an important turning point for climate science and policy. Supported by governments around the globe, the Intergovernmental Panel on Climate Change (IPCC) was founded under the roof of the World Meteorological Organization (WMO) and the United Nations Environment Program (UNEP) as a unique science-policy interface. The panel, i.e. participating governments, tasked a set of elected scientists to assess the state of climate science in dedicated reports, i.e. to review and synthesize scientific information relevant to understanding the scientific basis of climate change and of its risk, its environmental, political, and economic impacts and possible response options (see https://www.ipcc.ch/reports/). The latest report is from 2021 (IPCC, 2021).

These assessments follow strict principles and procedures (see https://www.ipcc.ch/site/assets/uploads/2018/09/ipcc-principles.pdf and https://www.ipcc.ch/site/assets/uploads/2018/09/ipcc-principles-appendix-a-final.pdf) to ensure policy relevance without being policy prescriptive. Hundreds of scientists and other experts contribute to the assessment in diverse author teams from a wide range of disciplines including climate physics, engineering, economics, geography, political science, psychology, sociology or urban science and from different world regions to ensure balanced findings. Review is another critical element of IPCC reports. Authors have to respond to tens of thousands of submitted comments by experts and governments in two rounds of review. Important for the dignity of IPCC assessments in the political sphere is the formal acceptance of the reports by the 195 member countries and the line-by-line approval of the summary for policymakers (Bolin, 2007; Edenhofer & Minx, 2014; Field & Barros, 2015; Mach, Freeman, Mastrandrea, & Field, 2016).

IPCC has been designed and used as the prime scientific input to international climate diplomacy under the United Nations Framework Convention on Climate Change – and as such contributed to international climate agreements – most importantly, the Kyoto Protocol and the Paris Agreement. Meanwhile, climate policy has become an integral part of most national policy programs. These programs include political actions that governments take to achieve the goal of limiting climate change and its consequences (see Yang, Huang, & Su, 2020).

In its summary for policymakers, the Climate change 2014 synthesis report (IPCC, 2014) states that "human influence on the climate system is clear, and recent anthropogenic emissions of greenhouse gases are the highest in history. Recent climate changes have had widespread impacts on human and natural systems" (p. 2). A recent study found that detectable and attributable climate impacts are documented in tens of thousands of scientific studies affecting 80% of the world's land area, where 85% of the world population resides (Callaghan et al., 2021). As such, it is unsurprising that the topic of climate change has become a hot topic in political and public debates and now features widely on political agendas across many different fields.

In this study, we deal with the question of how research on climate change and policy are connected. According to Yin, Gao, Jones, and Wang (2021), the systematic understanding of the connection between science and policy is still limited, since reliable data are missing on a global scale. In 2019, however, the new Overton database of policy documents was released including links to research papers that are cited by policy documents. Yang et al. (2020) define policy documents in this context as "'carriers' of policies … [that] provide a channel through which policy science researchers can study the main contents of policies, policymaking processes and policy instruments". Using Overton data, Yin et al. (2021) analyzed the connection between science and policy with respect to COVID-19. They found that "many policy documents in the COVID-19 pandemic substantially access recent, peer-reviewed, and high-impact science. And policy documents that cite science are especially highly cited within the policy domain. At the same time, there is a heterogeneity in the use of science across policy-making institutions. The tendency for policy documents to cite science appears mostly concentrated within intergovernmental organizations (IGOs), such as the World Health Organization (WHO), and much less so in national governments, which consume science largely indirectly through the IGOs" (p. 128).

Impact measurement of scientific papers on the policy area is part of a new branch in scientometrics: measurement of societal impact (Bornmann, 2016). Whereas science impact measurements of papers were restricted to citation analyses (using Web of Science, WoS, or Scopus data) until recently, societal impact measurements are focused on impact analyses of papers on other parts of society than science (Tahamtan & Bornmann, 2020). One part of the society is of special interest in this respect: the policy area. The policy area is permanently required to find answers on certain societal demands (such as COVID-19 or climate change). Since science permanently produces research results that can (and should) be used in the response to these demands, it is interesting to know, whether and to what extent this happens. Fang, Dudek, Noyons, and Costas (in press) defines the term 'policy impact' in this respect as impact that "tells the story of how research outputs provide concrete evidence to support policy-making processes, which can be reflected by the references to research outputs in policy documents". The use of research findings in the policy-making process is denoted as evidence-based policy-making (Black & Donald, 2001) or science-based policy-making (Pedersen, 2014). OPENing UP (2016) regards "informing policy and influencing decisions … as one of the most notable effects of scientific research" (p. 24).

## 2  Overview of studies on policy impact

The overview of studies dealing with the use of scientific information/publications in policy making by Vilkins and Grant (2017) reveals that a number of studies exists that are based on interviews and surveys (with policymakers). These studies show, e.g., that the use of scientific publications in policy documents seems to depend on organizational culture and perspectives towards their use. Furthermore, some policy areas (such as information technology) use scientific information more frequently than others (e.g., immigration or justice). The use of scientific information in policy might be distinguished according to three stylized purposes: "'instrumental' use is direct and measurable for policy; 'conceptual' use … [is] indirect but rather affects thinking over a longer period of time; 'symbolic' use is when specific findings are selected for rhetorical or political argument" (Vilkins & Grant, 2017, p. 1683). Sources of scientific information preferred by policymakers are the internet, meetings, and emailing colleagues. Yang et al. (2020) reviewed some studies that have analyzed networks of policymaking institutions to gain insights into their relationships. These studies focused on policymaking organizations' networks, public service organizations' networks, and policy collaboration networks.

In the area of altmetrics research, a recent overview of studies on measuring policy impact using altmetric data can be found in Fang et al. (in press) and Yang et al. (2020). A number of studies has used policy impact data from Altmetric (https://www.altmetric.com) or PlumX (https://plumanalytics.com) (see Arthur, 2016; Michalek et al., 2017). Very recent studies used Overton data (e.g., Yin et al., 2021). In the following, we summarize some of these policy impact studies chronologically. One of the first studies in this new altmetrics area was published by Bornmann, Haunschild, and Marx (2016) using an extensive publication set of climate change papers. The authors were interested in the question of how intensively policy documents have cited science publications. Although climate change is an ongoing policy topic worldwide, they found that only 1.2% out of 191,276 papers on climate change in the dataset have at least one policy citation (using data from Altmetric). The results of Bornmann et al. (2016) revealed that review papers were more frequently cited in policy documents than articles. In order to investigate whether the percentage of 1.2% can be thought of as high or low, two of the authors investigated the percentage of papers indexed in the WoS that are mentioned in policy-related documents (Haunschild & Bornmann, 2017). They found that less than 0.5% are mentioned at least once. Thus, the results show that although only 1.2% of climate change papers were relevant for policy documents, this percentage is substantially higher than the percentage among all papers from the database.

Vilkins and Grant (2017) did not use data from Altmetric or PlumX for their empirical study, but used publications from policy-focused Australian Government departments. The authors were interested in the research and reference practices of Australian policymakers. The study is based on 4,649 cited references in 80 government publications from eight departments. They found that mostly peer-reviewed journal articles, federal government reports, and Australian business information have been cited. The study also revealed "a possible increased chance for academic research to be cited if it was open access. Despite criticisms of citation analysis, at least in the field of research utilisation we cannot solely rely on interview or survey data, as cited evidence use differs from reported evidence use" (Vilkins & Grant, 2017, p. 1681).

Tattersall and Carroll (2018) used Altmetric policy documents data to investigate policy impact of papers published by authors at the University of Sheffield. They found that 0.65% of the papers were cited by at least one policy document. This percentage is slightly higher than that mentioned by Haunschild and Bornmann (2017) for the WoS database. The field-specific policy-impact analysis revealed that "the research topics with the greatest policy impact are medicine, dentistry, and health, followed by social science and pure science" (Tattersall & Carroll, 2018). In a more recent study, Yang et al. (2020) used the Chinese database iPolicy that includes policy documents issued by the Chinese government since 1949. The authors used the data to construct networks of policy-making ministries and government departments. They were interested in identifying core policymakers in China and possible changes of their positions in the networks. Yang et al. (2020) present 15 ministries in China with the highest eigenvector centrality as core government ministries in the policy networks.

Fang, Costas, Tian, Wang, and Wouters (2020) focused on hot research topics reflected by citations in policy documents (using Altmetric.com data). The study is based on more than 10 million WoS papers published in various disciplines. The authors identified the hot topics in various broad disciplines. For example, they found that infectious diseases were typically of concern to policy-makers, but also topics that focus on industry and finance as well as child and education. In addition, "potential health-threatening environment problems (e.g., 'ambient air pollution', 'environmental tobacco smoke', 'climate change', etc.) drew high levels of attention from policy-makers too" (Fang et al., 2020).

Hicks and Isett (2020) published a case study that investigated the policy impact of papers published in the area of quantitative studies of science. The authors speculated that many papers in this area have limited policy impact, but some papers such as the papers selected for their case study received a lot of policy impact. Hicks and Isett (2020) explain in detail the policy impact of the selected papers. For example, the authors selected the well-known study by Mansfield (1991, 1998) that estimated the social rate of return to public research spending. Hicks and Isett (2020) describe the diverse policy impact reached by this paper using several sources.

In the most recent study, Pinheiro, Vignola-Gagné, and Campbell (2021) used publication data from Framework Programmes (FPs) for Research and Technological Development. The authors investigated the relationship of cross-disciplinarity on the paper level and policy impact measured by policy citation data from the Overton database. Pinheiro et al. (2021) conclude as follows: "Our approach enables testing in a general way the assumption underlying many funding programs, namely that cross-disciplinary research will increase the policy relevance of research outcomes. Findings suggest that research assessments could benefit from measuring uptake in policy-related literature, following additional characterization of the Overton database; of the science-policy interactions it captures; and of the contribution of these interactions within the larger policymaking process" (p. 616).

# 3    Dataset used

For many years, policy documents' and policy citations' data were aggregated only by the companies Altmetric and PlumX. Recently, however, the Overton database (see https://www.overton.io) was launched with the goal of becoming the largest database of policy documents and citations (Szomszor & Adie, 2022). In Overton, policy documents are defined "very broadly as documents written primarily for or by policymakers" (see http://help.overton.io/en/articles/3823271-what-s-your-definition-of-a-policy-document). Overton includes documents from governments, think tanks (i.e., research institutions that perform research and advocacy in climate change), non-governmental organizations (NGOs) and intergovernmental organizations (IGOs, i.e., organizations that are composed of states) (see http://help.overton.io/en/articles/5062448-which-publications-does-overton-collect). The database includes not only various bibliographic information on policy documents (e.g., title and appearance), but also the citation links that exist between policy and science as well as among the policy documents in the database themselves. The citation relations are identified by Overton by using text-mining methods. According to Yin et al. (2021), the Overton database "includes all major economies and large population centers, with a notable exception of mainland China" (p. 128). The database is updated on a weekly basis. In December 2020, the database includes 799,716 policy documents with citation relations to either other policy documents or scientific papers in 66 different languages from 168 countries (including the European Union and IGOs) and more than 1250 different policy sources.

Yin et al. (2021) studied the reliability of the science-policy citations in the Overton database, by comparing them with the citation links provided by the Microsoft Academic Graph database (see https://academic.microsoft.com/home). The results show that "although the two datasets are collected for different purposes using different approaches and technologies, the measurements carried out independently across the two datasets show remarkable consistencies" (p. SI). Since the results by Yin et al. (2021) could confirm the reliability of the Overton data, we decided to use the data for the current study on climate change. Overton provided a snapshot (dated December 04, 2020) of their database to some of us (LB and RH). This snapshot has been imported into a local PostgreSQL database at the Max Planck Institute for Solid State Research (Stuttgart, Germany). After an analysis of publication dates of policy documents and consultation with Euan Adie (Overton), we excluded the policy documents with the publication dates '1970-01-01', '1970-01-02', and '2002-07-01' from our analysis because they were confirmed as 'dummy' publication dates by Euan Adie or contained many policy documents published later than the specified date.[1] We used PostgreSQL and R (R Core Team, 2019) commands including the R package 'tidyverse' (Wickham, 2017) for data analysis.

We searched in the fields 'title', 'translated title', and 'snippet' for climate-change-related terms in the Overton snapshot. We searched for 'climate change' and 'global warming' (note that both terms were truncated at both sides and a single arbitrary character was allowed instead of the white space between the words) to cover the bulk of policy documents that are related to climate change. The search strategy is based on keyword analyses in connection with search queries of previous climate change related papers (Bornmann et al., 2016; Haunschild & Bornmann, 2017). We found 10,846 policy documents that met the climate change search criteria out of 799,716 policy documents with any citation relation to a scientific paper or another policy document.

The Overton database includes links to scientific publications via digital object identifiers (DOIs) – "scholarly" references in Overton must have a DOI. There are 8,533,973 citation relations from 492,958 policy documents to 3,242,626 scientific papers. We used the SciTech Strategies' in-house version of Scopus containing 52.04 million items indexed as of

---

[1] See https://help.overton.io/article/why-am-i-seeing-unknown-date-instead-of-a-publication-date.

May 2020 and published between 1996 and 2019 as a database for scientific papers. 76.7% of these items have a DOI. We were able to match 2,071,085 DOIs cited in Overton to Scopus papers. Thus, nearly 4.98% of Scopus items with a DOI have been cited by policy documents indexed in the Overton database. This is substantially higher than the 1.12% mentioned in (Fang et al., 2020).

We used the journal metric CiteScore to measure the citation impact of journals (Teixeira da Silva, 2020). It is the mean number of citations for papers published in a journal. For the current study, CiteScore values were downloaded from https://www.scopus.com/sources.uri on November 10, 2020. The most current CiteScore values from 2019 were used for our analyses.

## 4   Results

### 4.1   Policy documents

This study is based on 10,846 climate change policy documents covered in the Overton database. This corresponds to 1.36% of all policy documents in the database. Figure 1 shows the distribution of the climate change policy documents across publication years. For a better interpretation of this distribution, we also included distributions for all policy documents in the Overton database and the papers on climate change in the Scopus database. The comparison of climate change with all policy documents reveals that the climate change policy documents reached a plateau in 2015 whereas all policy documents steadily increased until 2018. Since the scientific paper distribution also shows a steadily increasing trend, it seems that the discussion of climate change in the policy area reached its maximum several years ago (at least temporarily).

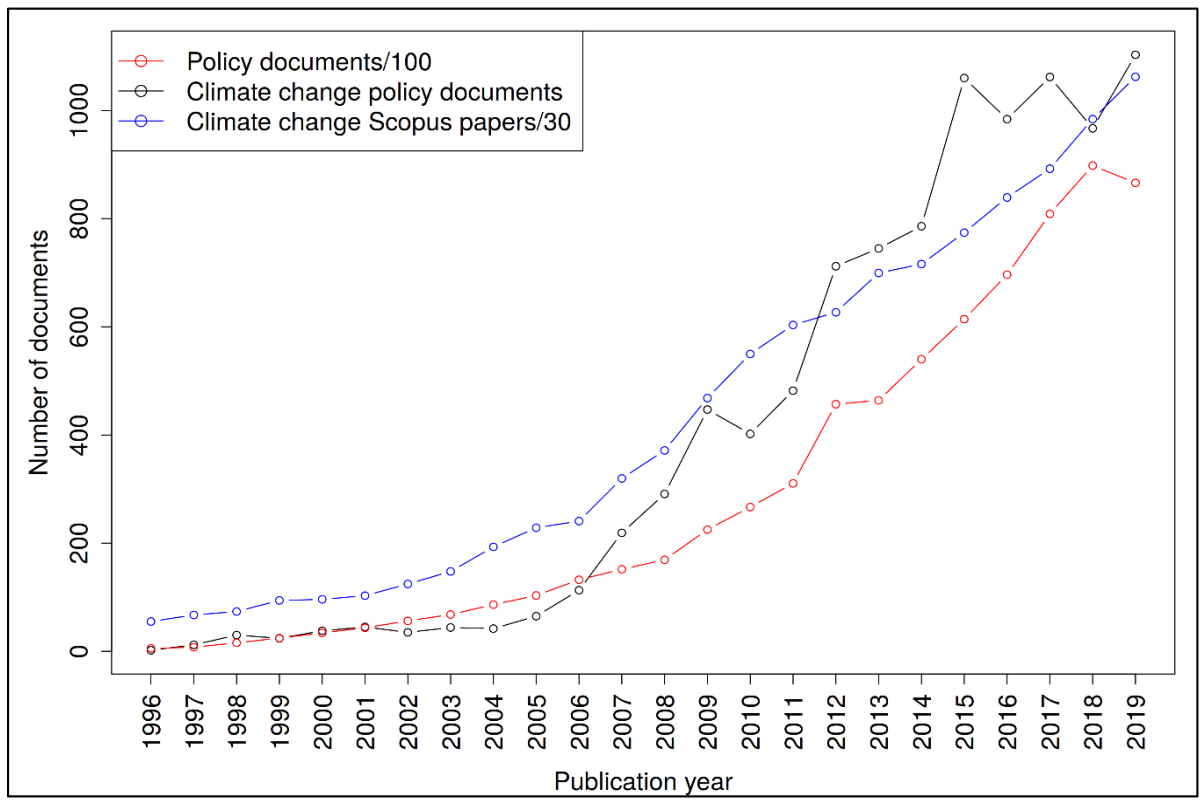

Figure 1. Annual number of policy documents and climate change policy documents

Policy documents can be published by various types of institutions. Based on the classification of these institution types used in Overton, Figure 2 shows the percentage of policy documents published by think tanks, governments, and IGOs. The comparison of climate change policy documents with all policy documents in Figure 2 reveals that climate change documents were published by think tanks and IGOs at higher than expected rates; fewer climate change documents were published by governments than expected.

This substantially lower share of climate documents issued by governments could be a reflection of their hesitance in dealing with the problem of climate change as documented in continued emissions growth (Friedlingstein et al., 2021; Lamb et al., 2021; Minx et al., 2021) as well as the gap between long-term ambition and short-term actions (Ou et al., 2021; United Nations Environment Programme, 2021). Another explanation could be that the share of documents from NGOs and IGOs is much higher in the field of climate – they might be particularly active in this space. IGOs, for example, may consider climate change as a problem of international coordination in nature.

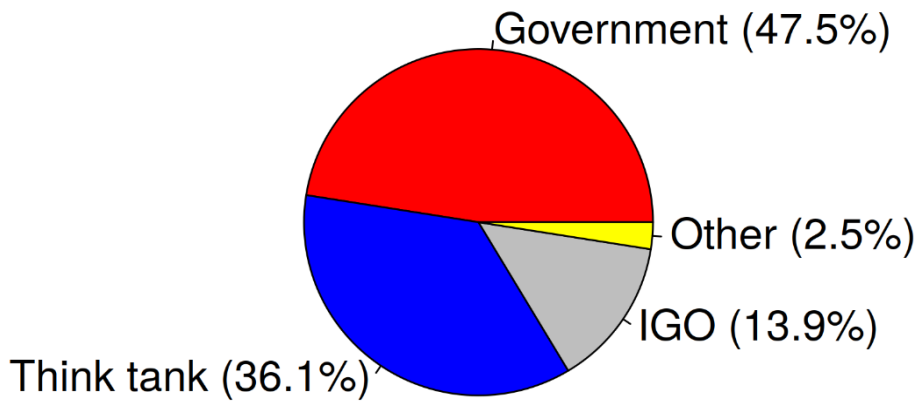

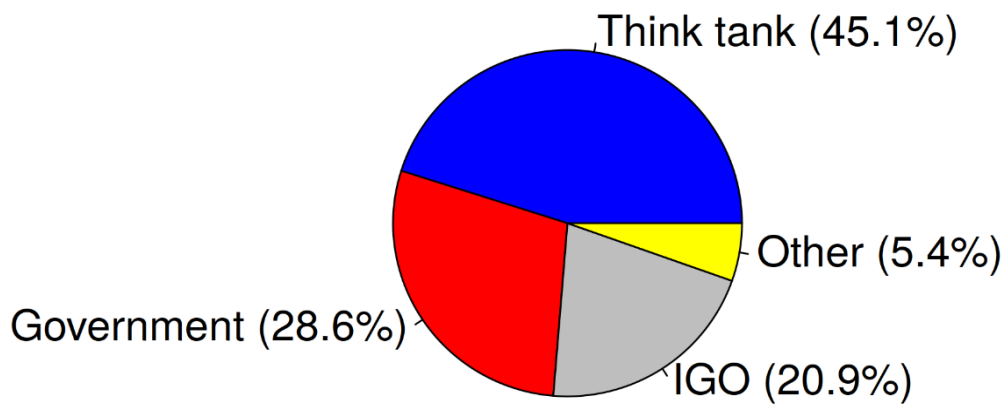

Figure 2. Percentage of (climate change) policy documents per sector

Figure 3 analyzes sectors publishing policy documents in more detail by considering single institutions. The figure shows the relationship for single institutions between number of policy documents and number of climate change policy documents. On the one hand, the results reveal those institutions (with high output) that are focused on climate change and those institutions that deal with climate change besides other topics. For example, due to its focus on a sector that is highly vulnerable to climate change, documents by the Food and Agriculture Organization (FAO) of the United Nations cover frequently the topic of climate change (please see the interactive version of Figure 3).

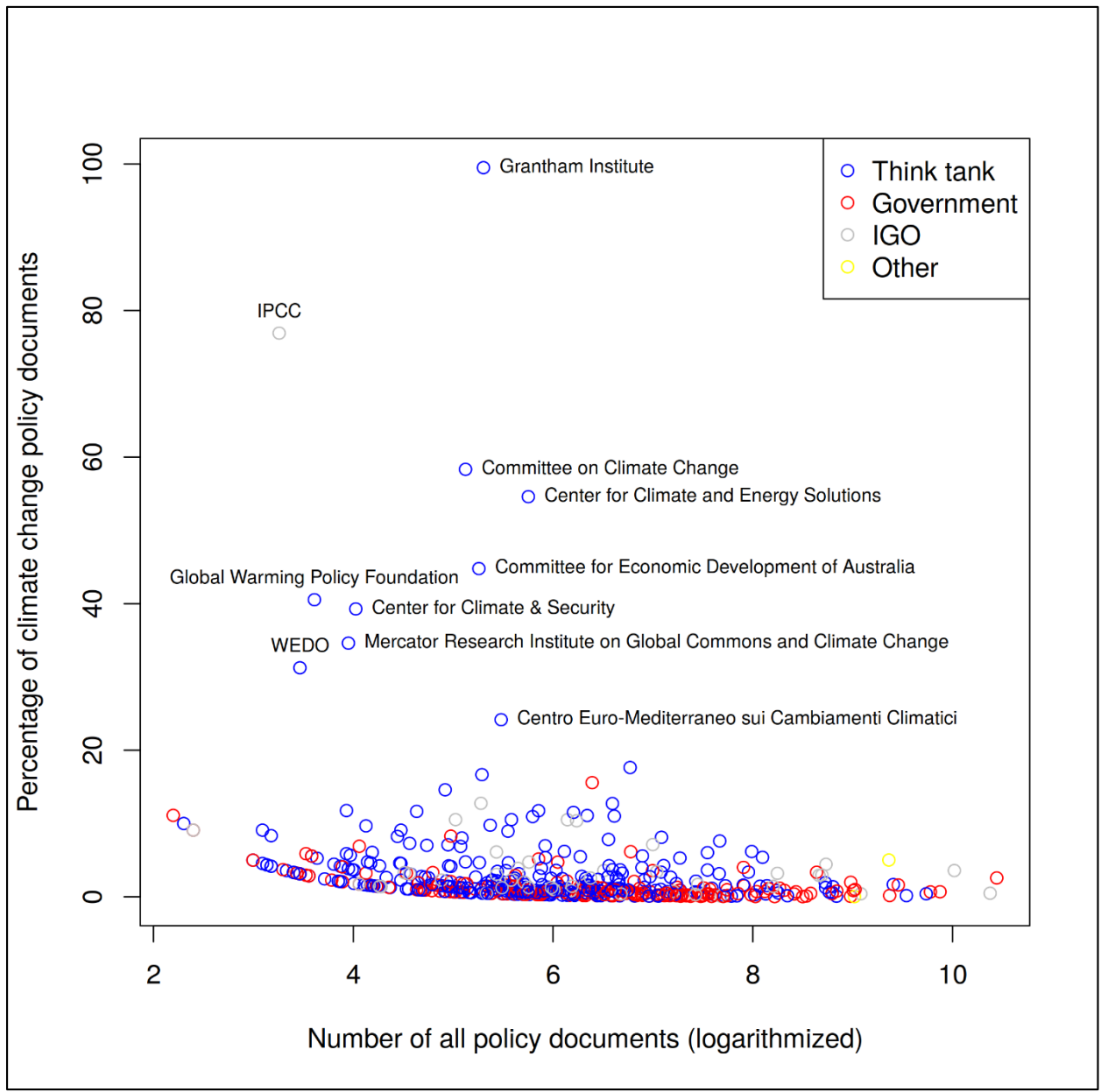

Figure 3. Correlation between the number of policy documents and the number of climate change policy documents on the institutional basis (Spearman rank correlation = 0.55; an interactive version can be viewed at: httpps://s.gwdg.de/Vtvdiw)

This is different in the field of health. Policy documents by the World Health Organization often do not cover climate change, even though this is starting to change now. This corresponds to the comparatively small share of publications in the field of medicine related to climate change research (Haunschild, Bornmann, & Marx, 2016) – even though there is a sizable and fast-growing number of research papers on climate and health in absolute terms (Berrang-Ford et al., 2021). On the other hand, the colors of the institutional dots in Figure 3 point out the relatively high number of think tanks and IGOs with a focus on climate change – of which some like the Global Warming Policy Foundation are alleged to focus on global warming misinformation and 'climate sceptic' contents (https://www.desmog.com/climate-disinformation-database/).

## 4.2 Papers cited in policy documents

In this section, we additionally consider the literature cited by climate change policy documents. We would like to know, for example, (1) whether these documents focus on recently published or older science literature and (2) the research institutions that seem to be very important for the policy area (since they were frequently cited). Figure 4 shows the document types of the publications cited by climate change policy documents. In order to facilitate the interpretation of the results, the results for all policy documents have been added. We have aggregated "article in progress" with "article". The type "other" contains empty document type entries, "abstract", and "missing". The results in the figure show that most policy documents reference "articles", followed by "reviews" and "conference papers". The other document types play a minor role. The referencing behavior seems rather similar in policy documents in general and in policy documents that are related to climate change.

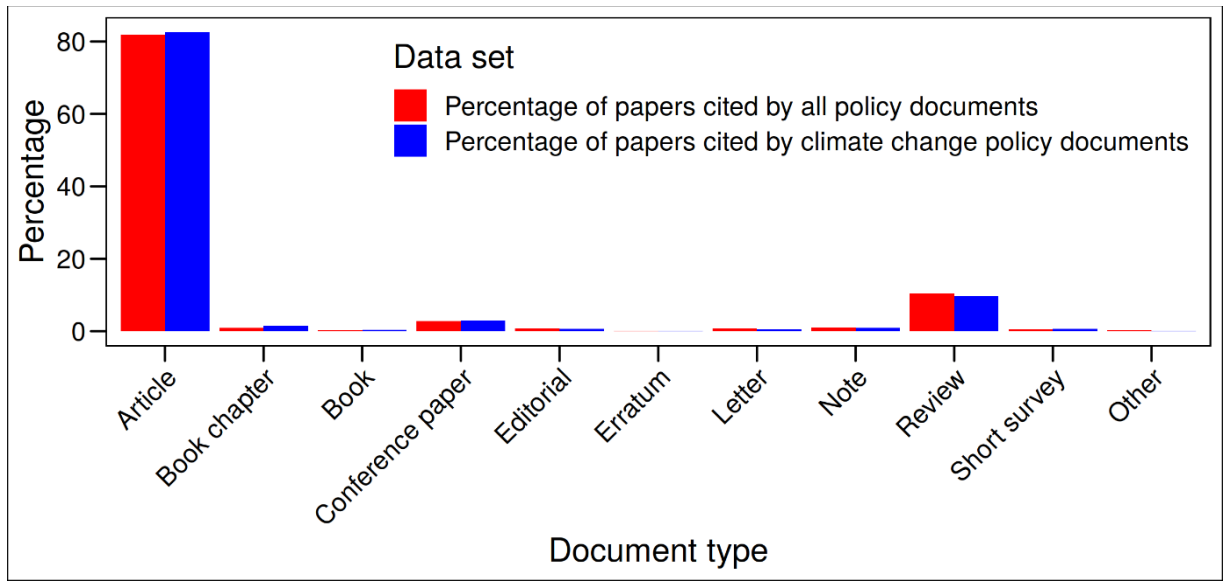

Figure 4. Document type of papers cited by policy documents (all policy documents and climate change policy documents)

Yin et al. (2021) found that "the COVID-19 policy frontier appears to be deeply grounded in extremely recent, peer-reviewed scientific insights" (p. 129). We expect there to be a similarly short time lag for climate change research on the one hand; but we can imagine a "classics" effect that certain foundational papers are referred to over and over again on the other hand (some of the policy documents might actually reiterate outdated findings/outliers as well). For scientific papers that cite other scientific papers, the results indicate a "classics" effect: if we look at cited references in papers, the average reference age is 13.1 years for all items in Scopus from 1996 to 2019. However, on average, climate change papers (published between 2010 and 2019) cite other scientific papers that are on average 9.7 years old. In this study, we also investigated the time between appearance of the policy document and its cited scientific papers. This difference is on average 5.8 years for climate change policy documents and 6.7 years for all policy documents. Both differences are significantly shorter than the average references ages in scientific papers and correspond to the results by Yin et al. (2021).

Figure 5 shows the proportions of accumulated citations of scientific papers in climate change policy documents over time. These proportions are compared with the proportions in all policy documents. We expected that climate change policy documents cite more recently published papers than other policy documents because of the great societal relevance of the topic. The results in Figure 5 show that this is indeed the case: the distribution for climate change policy documents increases faster than the distribution that refers to all policy documents. Yin et al. (2021) found a similar result for COVID-19 policy documents – another topic with high societal relevance.

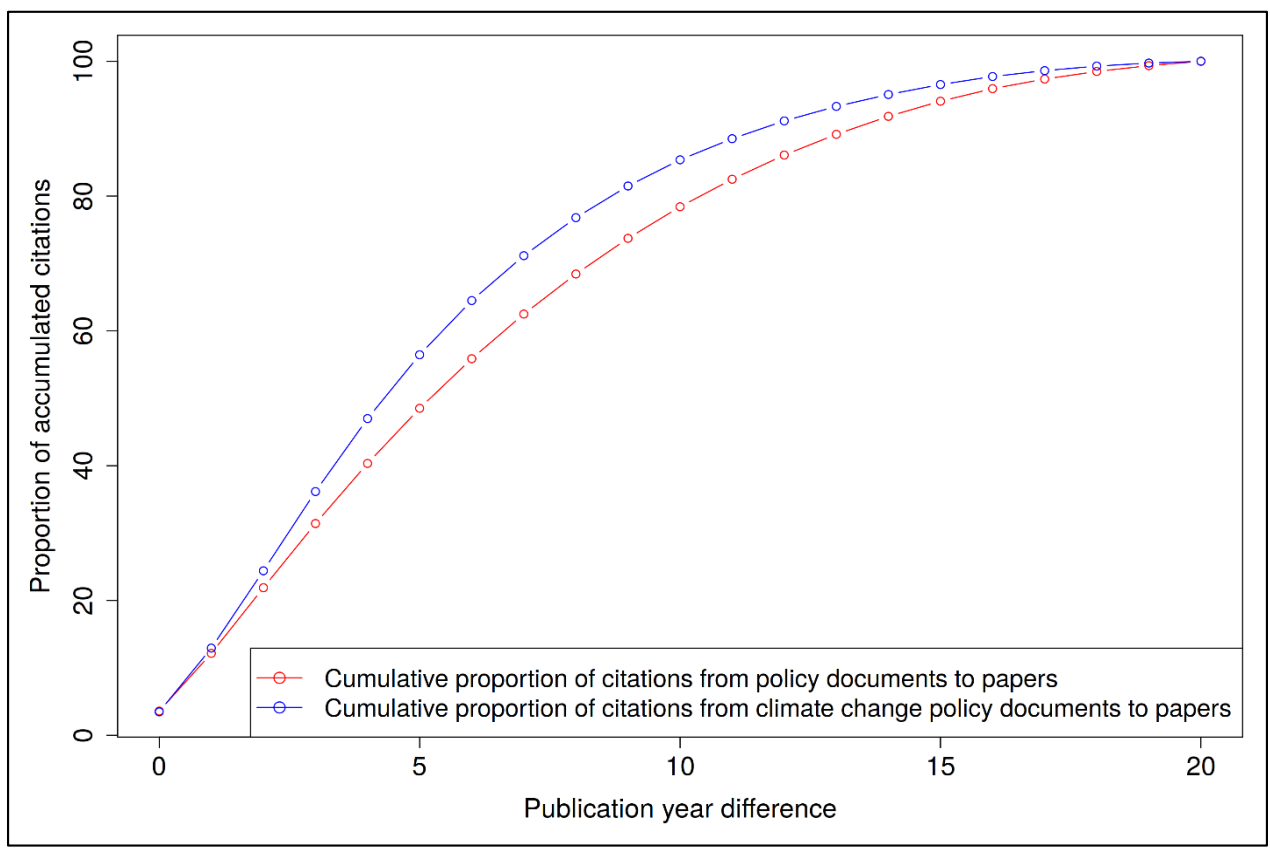

Figure 5. Proportion of accumulated citations of scientific papers in (climate change) policy documents over time. The publication year differences are the time between publication year of the policy document and publication year of the scientific paper.

We expected that policy documents preferentially cite papers published in reputable journals. The most valuable papers can be expected to be published in these journals. The results by Yin et al. (2021) show, for example, that "COVID-19 policy documents disproportionately reference peer-reviewed insights, drawing especially heavily on top medical journals, both general (such as Lancet) and specialized (such as Clinical Infectious Diseases)" (p. 129). In this study, we used CiteScore as the indicator for measuring reputation. Figure 6 shows the correlation between number of policy document citations received by papers in various scientific journals and the CiteScore of these journals. With a Spearman rank correlation coefficient of 0.24 (on the journal level), the relationship between journal reputation and policy citations is quite low.

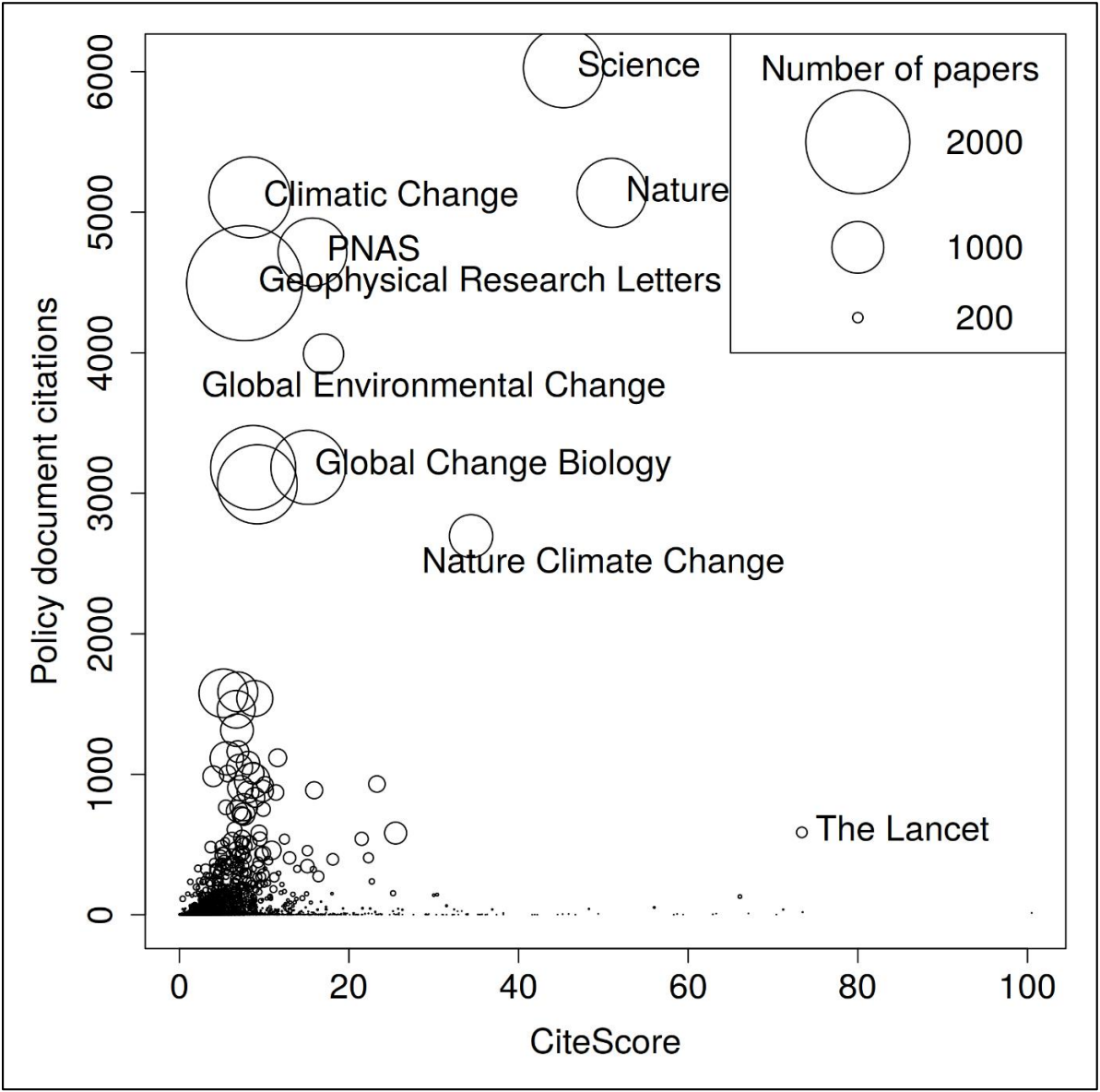

Figure 6. Correlation between number of climate change policy document citations received by papers in various scientific journals and number of Scopus papers published in these journals (Spearman rank correlation = 0.24; an interactive version can be viewed at: https://s.gwdg.de/k9Wp07)

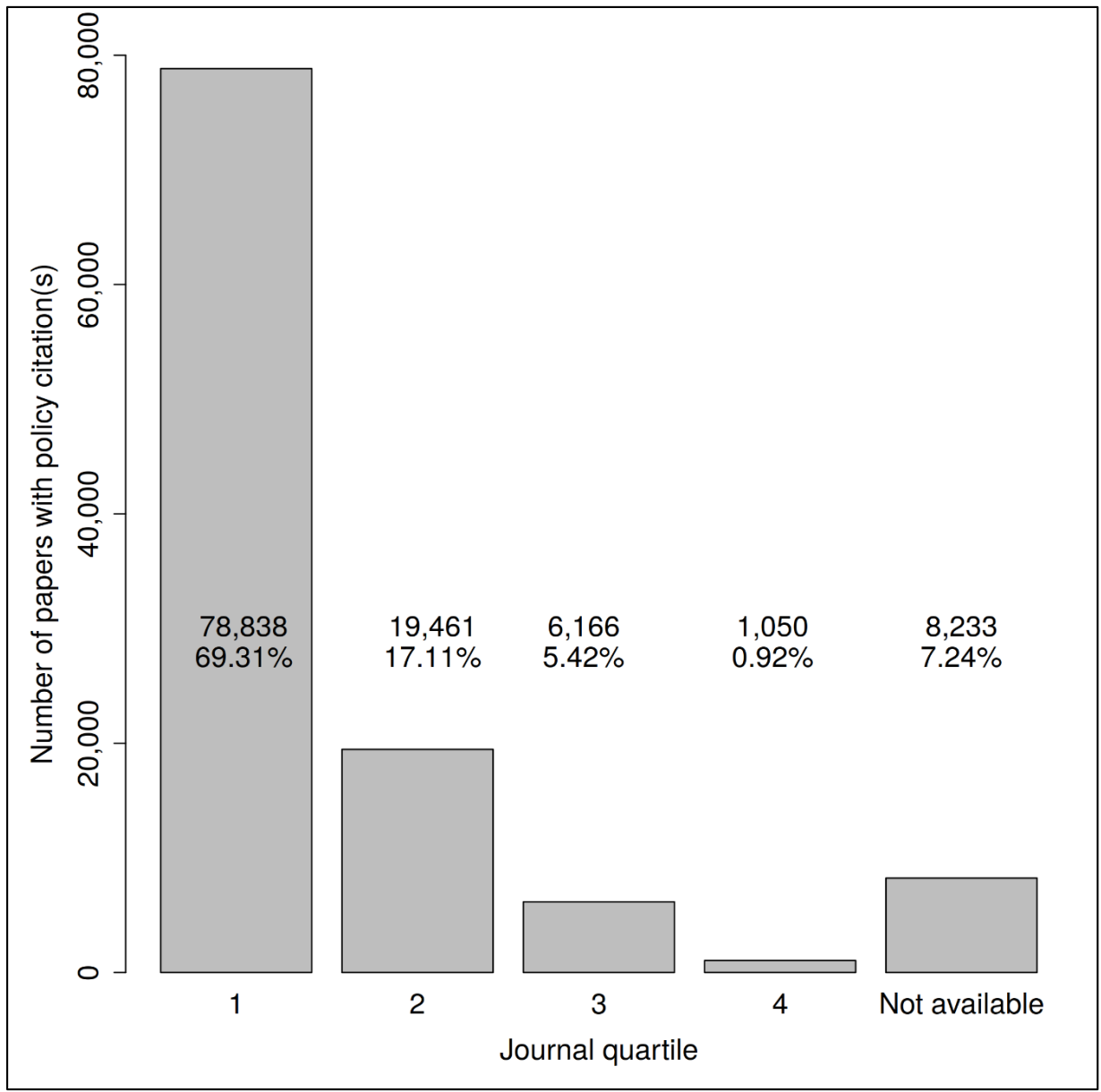

Figure 7. Number of papers with at least one policy citation grouped by CiteScore quartile of the journal. In the first journal quartile, e.g., are those journals that belong to the 25% of the journals with the highest CiteScore in their subject areas. For about 7% of the journals, a CiteScore was not available.

One reason for the low correlation might be that Citescore values at the top of the distribution are very spread out. If one were to use journal ranks rather than using Citescore, the coefficient would likely be much higher. In fact, this is the argument made in Figure 7. We found that scientific literature cited in policy documents is frequently published in high-impact journals: 69.31% of the papers with at least one policy citation were published in first-quartile journals. Thus, one can expect that policy citations of scientific papers correlate with citations of these papers in the scientific literature.

The results by Yin et al. (2021) for COVID-19 policy documents show that "the coronavirus research used by policy-makers aligns with what scientists heavily engage with themselves" (p. 129). In this study, the Spearman rank correlation coefficient between Scopus citations and policy citations of papers (n= 2,071,085) that were cited by policy documents at least once is 0.16. The correlation coefficient is slightly higher (0.20) between Scopus citations and policy citations of papers (n= 102,372) that were cited by climate change policy documents at least once. However, climate change papers that are cited in climate change policy documents received significantly more citations (between 3.3 and 5.6 times) on average than climate change papers that are not cited in these documents (see Figure 8).

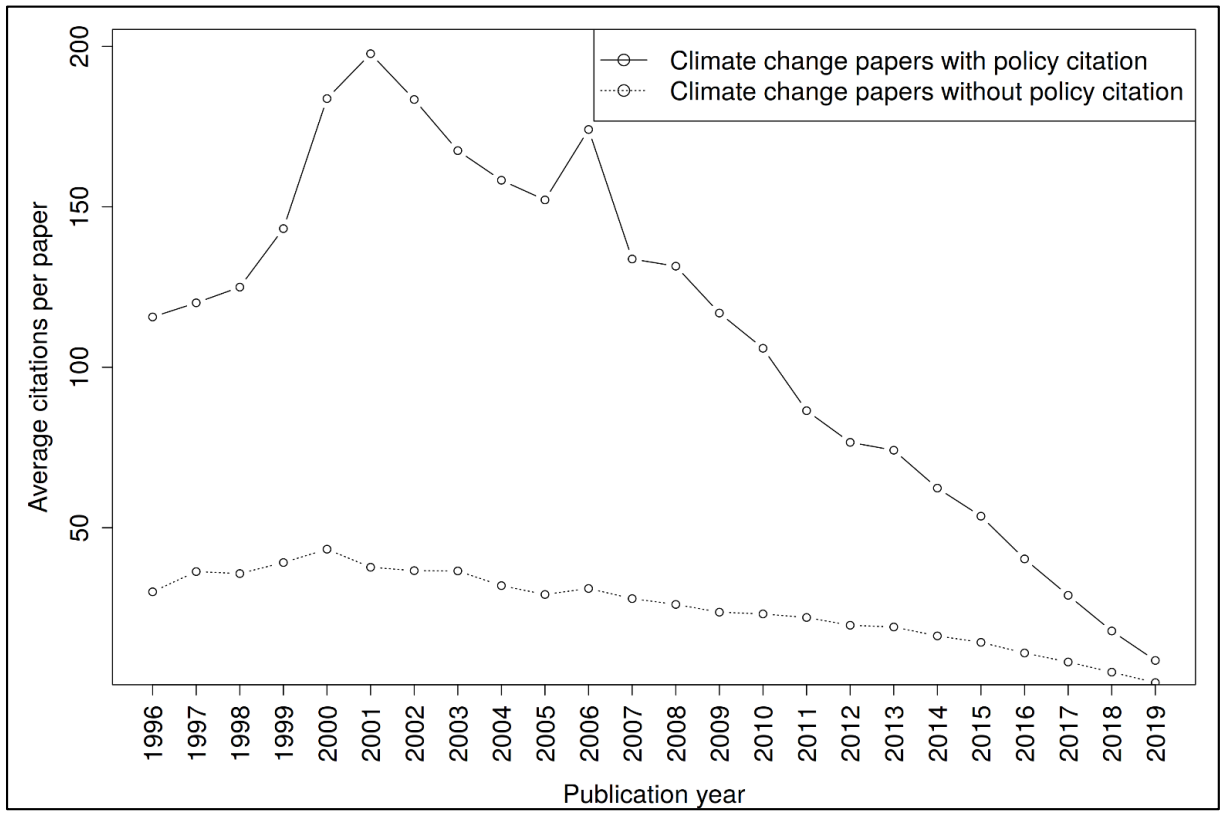

Figure 8. Average citations per paper of climate change papers in the scientific literature that are cited in policy documents (solid lines) or not (dotted lines).

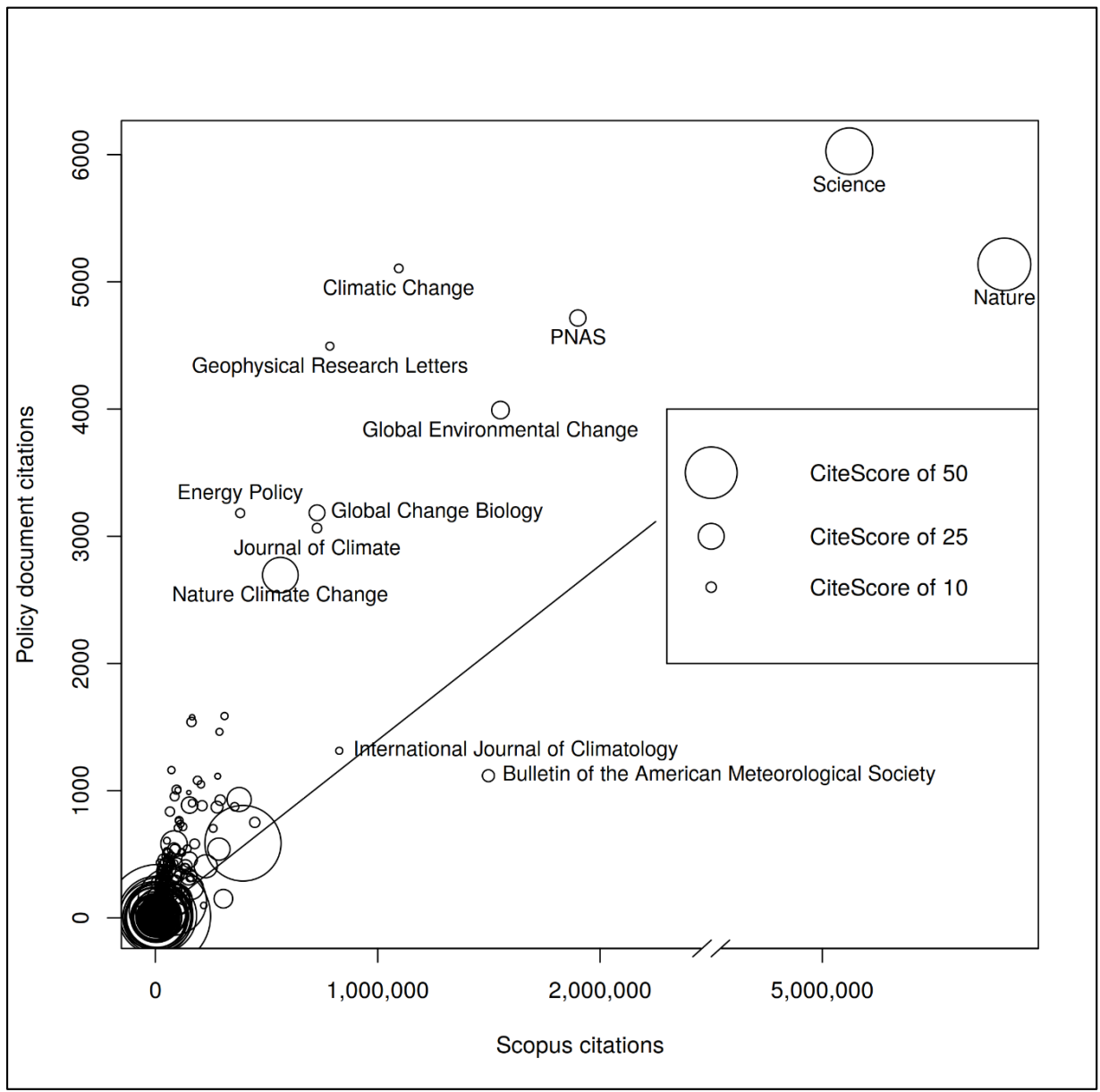

Figure 9. Correlation between the journal-based number of climate change policy document citations and Scopus citations. The size of the circles reflects the CiteScore of the journals (Spearman rank correlation = 0.81; an interactive version can be viewed at: https://s.gwdg.de/4weLvb)

Figure 9 includes the journal perspective to show the correlation between the number of climate change policy document citations and Scopus citations. The Spearman rank correlation between both citation counts is high at 0.81. The results in the figure point out that some journals receive more policy citations than can be expected based on science citations such as *Climatic Change* and *Nature Climate Change*. These climate change specific journals have emerged more recently. We speculate that the scientific communities of some highly specialized research topics are comparatively small, thereby limiting the mean number of citations per paper. *Nature* and *Science* papers received many citations in both areas of science and policy.

The journal analyses in the previous figures could not reveal the field-specific orientation of the papers cited in climate change policy documents. The journals that are labeled in the figures are mostly multi-disciplinary journals such as *Science* or *Nature* or are directly related to climate change. In order to explore the fields in which papers cited in climate change policy documents were published, we produced so called overlay maps that are presented in Figure 10. The overlay maps were created using the global mapping process outlined in Boyack and Klavans (2019). Here, clustering was done on 46.14 million Scopus-indexed documents (1996-2019) and 27.23 million non-indexed documents cited at least twice with over 1.1 billion citation links using the Leiden algorithm (Traag, Waltman, & van Eck, 2019). Graph layout was then done on the resulting 104,677 clusters using OpenOrd/DrL (Martin, Brown, Klavans, & Boyack, 2011) and cluster-level relatedness based on the bm25 text relevance measure, which has been shown to produce better clustering than a simple tf-idf measure (Boyack et al., 2011; Sparck Jones, Walker, & Robertson, 2000a, 2000b).

Over 11% of Scopus-indexed documents were not included in clusters or the map because they had no references and were not cited. Each cluster is represented as a dot on the map and was assigned to its dominant field (and colored) using the journal-to-field assignments from the so-called UCSD map of science (Börner et al., 2012). Clusters with similar topical content are close to each other on the map. Aggregations of clusters can be perceived as discipline-level structures; local areas that contain clusters of many colors are multidisciplinary. Although dot sizes for overlays are based on the number of documents matching overlay criteria, the intent is to provide a qualitative (gestalt) visual view of the data, e.g. to show where result sets are concentrated or if they are evenly spread throughout the map.

Figure 10 shows four maps for comparison: (1) All papers from Scopus, (2) Climate change papers in total, (3) Climate change papers with at least one policy citation, and (4) Papers with at least one policy citation. The maps focus on all papers in the Scopus database or are restricted to papers with climate change relatedness and/or policy citation. Comparing map (2) with map (3), for example, one can see that there are areas with climate change papers (such as computer science, pink in map 2) that are not well cited by climate change policy documents – there is far less pink in map 3 than in map 2.

Similar to all papers from the Scopus database shown in map (1) of Figure 10, papers with at least one policy citation extend across all scientific fields [see map (4) of Figure 10]. However, some major fields appear less pronounced in map 4: in particular chemistry, physics, computer sciences, and engineering. Biology, disease sciences, and health sciences are accentuated quite similar, indicating that in general these fields are more policy relevant. The fields of climate change papers in map 2 of Figure 10 are concentrated in biology, earth sciences, engineering, disease sciences, and physics (less pronounced). Climate change papers with at least one policy citation [see map 3 of Figure 10] show a field-specific pattern quite similar to the overall climate change policy papers in map 2. It seems that politics does not have a specific field, but reflects the field-specific orientation of climate change research.

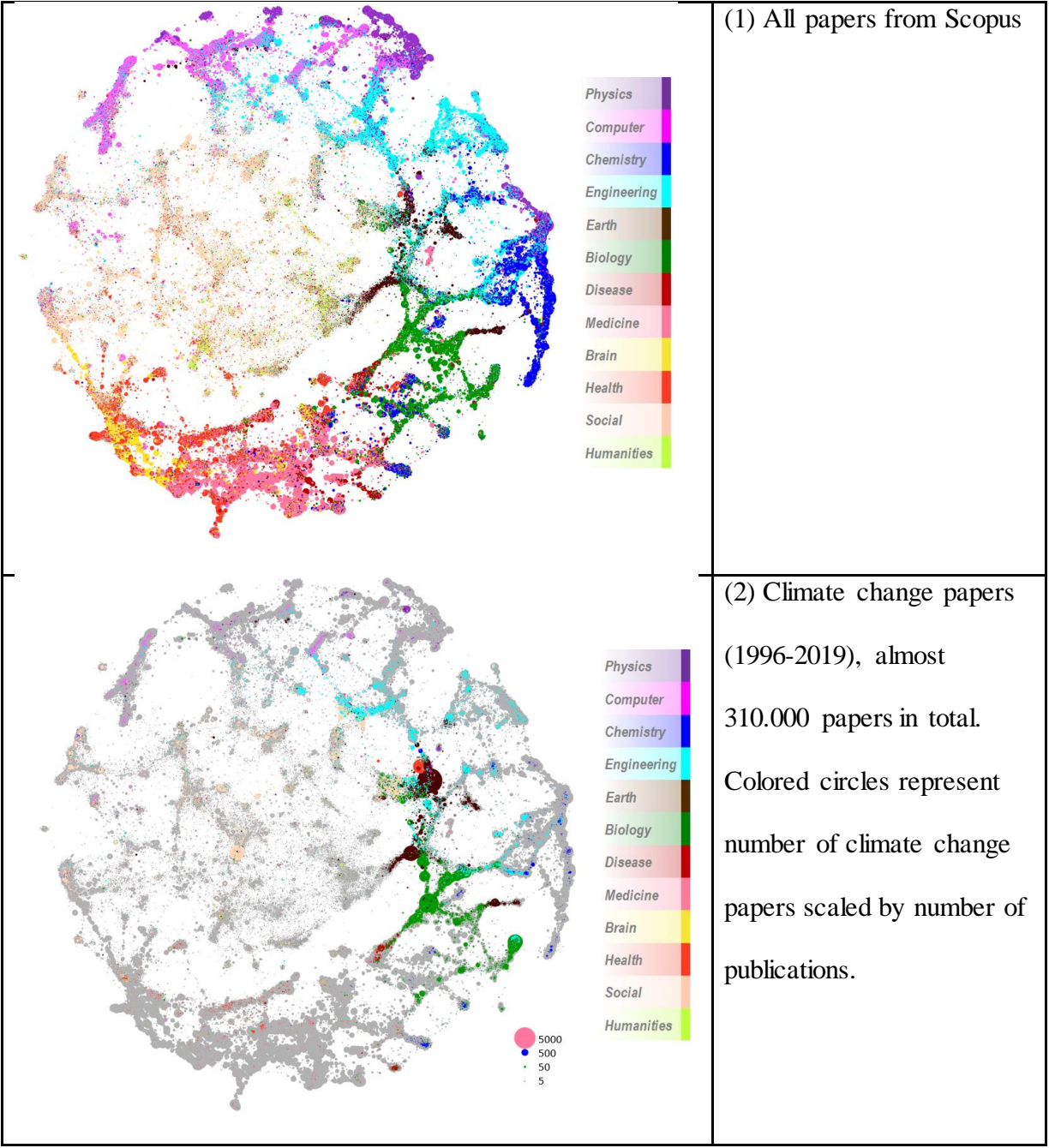

(1) All papers from Scopus

(2) Climate change papers (1996-2019), almost 310.000 papers in total. Colored circles represent number of climate change papers scaled by number of publications.

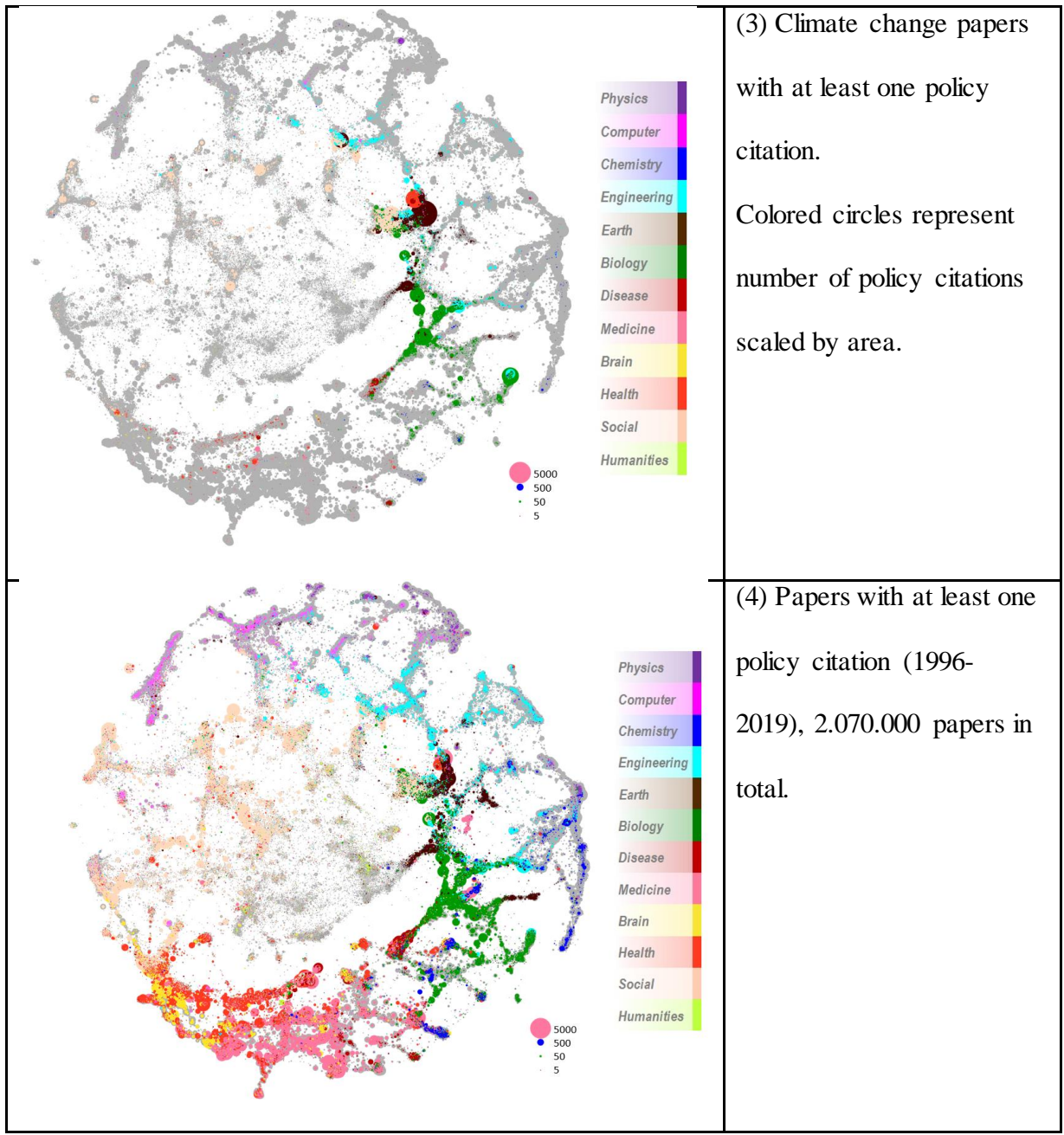

Figure 10. Overlay maps visualizing field-specific clusters of papers (based on citation relations). The maps include (1) all papers, (2) climate change papers, (3) climate change papers with at least one policy citation, (4) all papers in Scopus with at least one policy citation.

For COVID-19 research, Yin et al. (2021) investigated the temporal shift of the literature cited in policy documents concerning the field-specific distribution (compared to the whole policy literature). Their results reveal "a clear shift from drawing primarily on the biomedical literature to citing economics, society, and other fields of study, which is consistent with overall shifts in policy focus" (Yin et al., 2021, p. 129). In this study, we also investigated whether there is a field-specific shift using the 27 high-level ASJC journal categories. Figure 11 shows the field-specific orientation of papers (with policy citations) over the entire period (1996-2019). There are some subtle shifts but the early years (2000-2010) suffer from small number effects relative to the most recent decade. Climate change policy documents cite different fields than the whole. The large shifts shown in Yin et al. (2021) aren't seen here, but COVID-19 is a rather unique situation where social concerns followed after the medical ones on a short time scale.

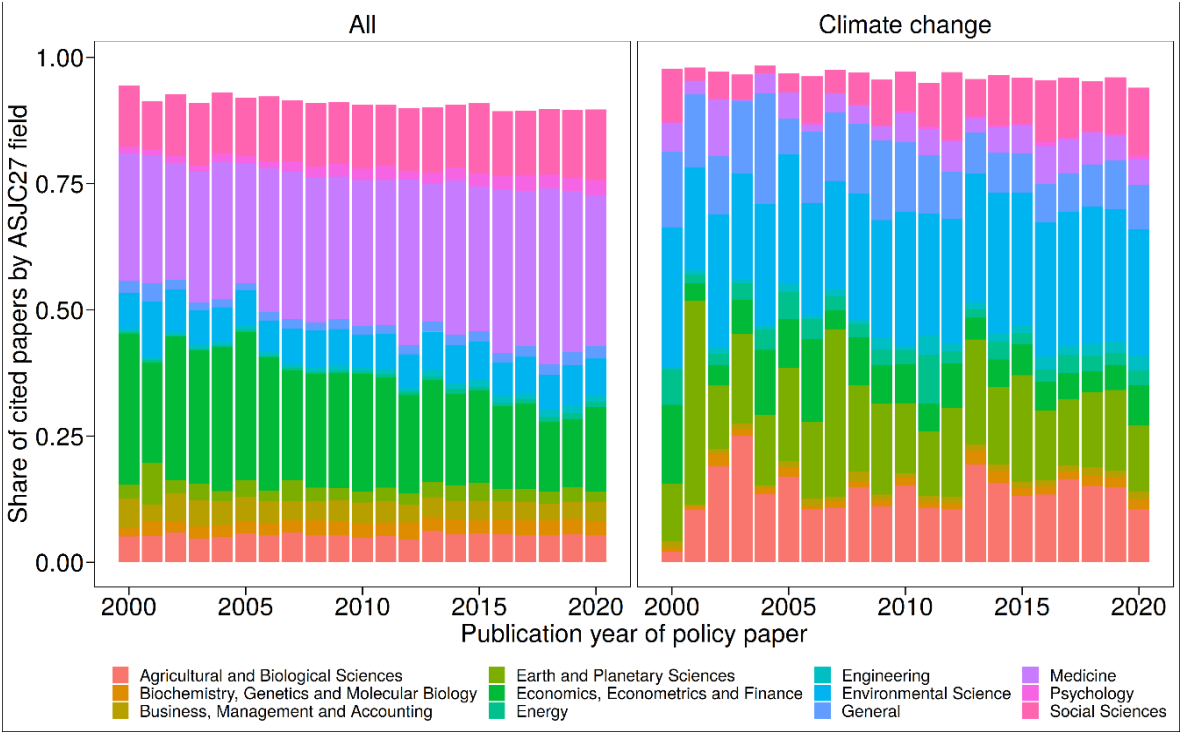

Figure 11. Fields of papers cited in policy documents (an interactive version with all ASJC27 fields is available at: https://s.gwdg.de/QTl1nm)

## 4.3 Scientific institutions and policy sources involved in political climate change discussions

In the final section of the empirical results, we focus on the scientific institutions and policy sources that are involved in the political climate change discussions. We are interested in the policy sources that are very active in political climate change discussions (and decisions) and science institutions that provide research results as inputs for the discussions. Table 1 shows the policy sources with the highest number of climate change policy documents. The table also reveals the number of scientific papers cited by these institutions and the number of climate change papers (the number in brackets is the number of policy documents citing the climate change papers). The results show that Publications Office of the European Union[2] and World Bank are the institutions with the most climate change policy documents. These cross-regional institutions are best-suited for dealing with global issues and thus are focused on major problems such as global warming.

Table 1. Policy sources with the highest number of policy documents (sorted by the number of policy documents). The table also reveals the number of scientific papers cited by these institutions and the number of climate change papers (the number in brackets is the number of policy documents citing the climate change papers).

| Policy source | Number of policy documents | Number of cited scientific papers | Number of cited climate change papers (number of citing policy documents) |
|---|---|---|---|
| Publications Office of the European Union | 33,096 | 279,161 | 13,601 (863) |
| World Health Organization | 23,993 | 268,872 | 2,213 (109) |
| World Bank | 19,476 | 169,787 | 15,402 (699) |

---

[2] According to Euan Adie (founder and director of Overton) the Publications Office of the European Union is a special case as it aggregates documents from many different EU agencies.

| | | | |
|---|---|---|---|
| National Bureau of Economic Research (NBER) | 16,766 | 119,655 | 1,124 (69) |
| Government Publishing Office | 14,910 | 231,200 | 10,003 (108) |
| IZA Institute of Labor Economics | 13,849 | 115,195 | 738 (22) |
| Government of Canada | 12,275 | 151,426 | 6,542 (196) |
| Analysis & Policy Observatory | 9,806 | 145,875 | 16,897 (508) |
| Centers for Disease Control and Prevention (CDC) | 9,179 | 156,840 | 458 (20) |

Table 2 focuses on policy sources that are rooted in climate change research. The results in the table reveal that IPCC is the source that referenced the largest number of papers. Considering the large amount of scientific information collected and presented in the various IPCC reports over many years, this is not surprising as the assessment of the scientific literature on climate change is its core mandate.

Table 2. Policy sources with the largest number of climate change papers cited. The table shows policy sources that cited more than 4.000 papers.

| Policy source | Number of papers cited |
|---|---|
| IPCC | 54,251 |
| Analysis & Policy Observatory | 38,529 |
| Publications Office of the European Union | 32,352 |
| World Bank | 28,156 |
| Government Publishing Office | 24,623 |
| United Nations Environment Programme | 20,029 |
| International Union for Conservation of Nature | 16,451 |
| Food and Agriculture Organization of the United Nations | 15,438 |
| Heartland Institute | 12,914 |
| Joint Research Centre | 11,164 |
| Arctic Council | 9,288 |
| Government of Canada | 9,165 |
| PBL Netherlands Environmental Assessment Agency | 7,151 |
| Committee for Economic Development of Australia | 6,325 |
| World Health Organization | 5,008 |
| GOV.UK | 4,725 |
| Umwelt Bundesamt | 4,478 |
| Asian Development Bank | 4,462 |
| New Zealand Ministry for Primary Industries | 4,422 |

We differentiated the results in Table 2 further by specifically looking at government, IGO, and think tank sources: we show policy sources in Table 3 that cite science for governments, IGOs, and think tanks. Yin et al. (2021) reveal the results of similar analyses based on COVID-19 datasets. The results show that governments and IGOs are of similar importance, both with regard to the overall number of policy documents and climate change related policy documents. The top ranked think tanks produced about half of the overall number of policy documents compared to the top ranked governmental organizations and IGOs. Their share of climate change research related documents is roughly the same.

Table 3. Types of policy sources most productive in publishing policy documents citing papers related to climate change research. The table differentiates between all documents of the sources citing these papers and documents focussing on climate change.

| All policy documents | | Climate change policy documents | |
|---|---|---|---|
| Policy source | Number of policy documents | Policy source | Number of policy documents |
| **Government** | | | |
| Publications Office of the European Union | 34,286 | Publications Office of the European Union | 881 |
| GOV.UK | 19,431 | Government of Canada | 206 |
| Government Publishing Office | 17,649 | Joint Research Centre | 189 |
| Government of Canada | 12,797 | European Parliamentary Research Service | 157 |
| Centers for Disease Control and Prevention (CDC) | 11,733 | GOV.UK | 132 |
| UK Parliament Select Committee Publications | 8,297 | Government Publishing Office | 117 |
| U.S. Government Accountability Office | 8,160 | Umwelt Bundesamt | 109 |
| European Parliamentary Research Service | 7,947 | PBL Netherlands Environmental Assessment Agency | 93 |
| City of New York | 7,934 | UK Parliament Select Committee Publications | 82 |
| UK Parliament Research Briefings | 6,910 | U.S. Government Accountability Office | 69 |
| **IGO** | | | |
| World Health Organization | 32,082 | World Bank | 803 |
| World Bank | 22,427 | Food and Agriculture Organization of the United Nations | 275 |
| International Monetary Fund | 8,795 | OECD | 176 |
| Food and Agriculture Organization of the United Nations | 6,204 | Inter-American Development Bank | 169 |
| OECD | 5,953 | World Health Organization | 155 |
| Inter-American Development Bank | 5,787 | Asian Development Bank | 122 |
| Asian Development Bank | 3,814 | International Organization for Migration | 78 |
| UNESCO | 3,658 | United Nations Environment Programme | 53 |
| Organization for Security and Co-operation in Europe | 2,164 | UNESCO | 52 |
| World Meterological Organization | 1,719 | Arctic Council | 49 |
| **Think tank** | | | |
| National Bureau of Economic Research (NBER) | 16,925 | Brookings Institution | 204 |
| IZA Institute of Labor Economics | 13,884 | Grantham Institute | 200 |
| Brookings Institution | 12,173 | Center for Strategic and International Studies | 182 |
| Gatestone Institute | 6,891 | Overseas Development Institute | 177 |
| Atlantic Council | 6,650 | Center for Climate and Energy Solutions | 172 |
| Foundation for Economic Education | 6,617 | International Union for Conservation of Nature | 163 |
| RAND Corporation | 6,415 | World Resources Institute | 154 |
| The Heritage Foundation | 6,175 | Acton Institute | 118 |
| Acton Institute | 6,163 | Center for American Progress | 114 |
| Tax Foundation | 4,217 | Committee on Climate Change | 98 |

Note. In this paper, we used the Overton classification for assigning institutions to governments, IGOs, and think tanks. We found that some institutions may be misclassified. For example, the Committee on Climate Change is categorized as think tank, but is a committee by the UK government to support climate policies.[3]

Table 4 is related to the cited institution side of the science-policy link: which science institutions received the most citations from policy documents? The table presents reputable institutions of climate change research or research units located at universities, with the

---

[3] According to Euan Adie the "think tank" category also includes NGOs and quasi-research institutions; the category currently includes things that don't fit the other categories.

University of East Anglia with its long-lasting tradition in climate change research and meteorology at the top.

Table 4. Scientific institutions with the most citations in policy documents. The table includes all institutions with more than 2000 policy citations

| Science institution | Number of papers cited |
|---|---|
| University of East Anglia | 3,917 |
| Wageningen University & Research | 3,892 |
| Potsdam Institute for Climate Impact Research | 3,828 |
| University of Oxford | 3,716 |
| National Center for Atmospheric Research | 3,602 |
| Stanford University | 3,475 |
| University of California, Berkeley | 3,142 |
| Columbia University | 2,993 |
| Harvard University | 2,956 |
| University of Washington | 2,925 |
| International Institute for Applied Systems Analysis | 2,868 |
| Met Office | 2,800 |
| ETH Zurich | 2,682 |
| Princeton University | 2,549 |
| Australian National University | 2,429 |
| Joint Research Centre | 2,308 |
| Utrecht University | 2,278 |
| University of British Columbia | 2,163 |
| VU Amsterdam | 2,155 |
| University College London | 2,087 |
| University of Queensland | 2,044 |

It is noteworthy that throughout Table 2 to Table 4, we find institutions that are alleged to focus on climate misinformation according to the Climate Disinformation Database (https://www.desmog.com/climate-disinformation-database/) like the Heartland Institute, the Foundation for Economic Education, the Heritage Foundation, and Acton Institute; those are very active publishers of policy documents. The Acton Institute also features among the most prolific think thanks publishing policy documents related to climate change. In the overall climate change dataset, we found 17 policy organizations that are listed in the Climate

Disinformation Database. The organizations produced 99 policy documents (that cited any Scopus paper) within our dataset; these documents cited 6507 Scopus papers. That is 1.4% of the policy documents and 4.1% of the cited Scopus papers in our dataset.

# 5 Discussion

The use of results and recommendations from research on climate change might be reflected in citations of scientific papers in policy documents. Studies analyzing the impact of research on policy belong to the area of societal impact measurements in scientometrics (Tahamtan & Bornmann, 2020). According to Vilkins and Grant (2017), "capturing this impact on policy has significant potential benefits, including showing the impact of research on real-world settings, and building a better case for support for researchers and institutions or even broader research directions" (p. 1682). For Yin et al. (2021) policy-science citations may occur "for different reasons … including (i) instrumental uses (knowledge directly applied to solve problems); (ii) conceptual uses (research influences or informs the way policymakers think); (iii) tactical uses (citing research to support or challenge an idea) among others, suggesting the need to understand the semantics of the policy science citations" (p. SI).

This study focusses on the connection of climate change research and policy. The study is based on data from the (new) Overton database including policy documents (10,846 climate change policy documents covered in the database) and their citations of scientific publications. With this study, we followed other studies using Overton data investigating links between policy and research (e.g., on COVID-19). Although the Overton database captures a large collection of policy documents, potential biases in coverage and data sample cannot be excluded (Yin et al., 2021). For example, the Overton providers will not have access to many governmental archives, and if they have access, it will be restricted to only a part of the existing documents. Other shortcomings of Overton are mentioned by Yang et al. (2020): "the metadata of such policy documents cannot reveal the semantic information contained in the policy process. At the same time, some policy documents have unstructured features, so attribute identification and labeling may be required".

Overton uses a very broad definition of policy documents, i.e. "documents primarily written by and for policy makers". The idea behind this is to cover not only text that documents the policy or legislation itself in the corpus, but also documents that were written to inform or influence decisions. Our analyses do not distinguish between those two fundamentally different classes of policy documents. Documents written for policymakers are often written with the purpose to inform or influence documents authored by policymakers and are as such fundamentally different from documents authored by scientists. Moreover, under this wide umbrella definition there are very different types of documents: scientific assessments by the scientific community, legislations, policy reports by IGOs and NGOs, policy briefs, speeches etc.

The different nature of these documents explains some of the results here. For example, it is the main purpose of scientific assessments as those by IPCC to assess the state of knowledge in climate change research and inform international climate diplomacy and national climate policy with robust evidence. In nature, these assessments are comprehensive reviews of the literature with tens of thousands of references. On the other hand, policy briefs are designed for communications and often deliberately strip out literature sources. The policy impact analysis in this study, therefore to some extent simply highlights different policy document types. Any interpretation of policy impact of research can only be undertaken based on such an important caveat.

In this study, we empirically targeted several aspects of the connection between climate change research and policy. Focusing on the time trend of this connection reveals that the discussion of climate change in policy seems to have had its peak some years ago. Although we suspect a lot of uncertainty related to the coverage of policy documents in Overton, there seems to be an impact of international climate policy cycles on policy document publication. We observe local peaks in climate policy documents around major decisions in international climate diplomacy. For example, we observe temporal peaks in policy documents around the failed Copenhagen Summit in 2009 and the Paris Agreement; there is a growth in policy documents from IPCC's Fifth Assessment in 2013/2014 with a peak in 2015 when the Paris Agreement was made. IPCC reports might play a particular role as they are usually released 2-3 years ahead of major international climate diplomacy events and could trigger substantial co-publication activities. In 2023, the first Global Stocktake on progress with the Paris Agreement is scheduled with IPCC AR6 being released during 2021 and 2022. We might thus expect to see increases in climate change policy documents and citations to the scientific literature in the 2-3 years following.

Various types of institutions publish policy documents. Our results point out that IGOs and think tanks – with a focus on climate change – have published more climate change policy documents than expected (this result may be partly driven by the biased coverage of the Overton database). The policy documents published by the different types of institutions have especially cited more recent publications. Since climate change is of great societal relevance worldwide, research activities are on a high level (compared to other topics) that can be picked up in a timely manner by the policy area. Although one might expect that policy and science impact correlate (what is relevant for the scientific discourse might be equally relevant for the policy discourse), we found the opposite: the correlation between policy citations and science citations and the correlation between policy citations and the impact factor of the journals publishing the papers are both low. Thus, it seems that both areas of society (science and policy) focus on different papers from climate change research. If the scientific discourse and the policy discourse are scarcely related in terms of citation counts, one might expect that they focus on different fields. Our results reveal, however, that this is not the case: climate change papers with at least one policy citation are concentrated on similar fields as all climate change papers (biology, earth sciences, engineering, and disease sciences). Since field differences scarcely exist between both publication sets of interest, it would be interesting to explore in future studies how the differences can be characterized by other means.

What are the policy sources that are very active in the political climate change discourse and which scientific institutions provide the necessary scientific information? Our results show that the Publication Offices of the European Union, World Health Organization, and World Bank have published the most climate change policy documents. Since climate change is a worldwide problem and demand, it comes as no surprise that these cross-regional institutions have the highest publication output. The relevant science institutions for policy sources are mostly institutions with high reputation in science – this might be in contrast to the low correlation between science and policy citations on the single paper level. On the institutional level, policy sources seem to trust scientific institutions being renowned for reputable research on climate change (e.g., the University of East Anglia).

In this study, we found differences – in terms of citations – between research outcomes that are relevant for the scientific discourse and outcomes that are important for the policy area. This discrepancy has been found also in other studies. One reason for the differences might be barriers to academic outcomes from policy institutions such as access to climate change publications (Vilkins & Grant, 2017). Another reason might be missing summaries of research results that are understandable for people outside academia. Bornmann and Marx (2014) recommend therefore that researchers should write assessment reports (such as the IPCC) summarizing "the status of the research on a certain subject … Societal impact is given when the content of a report is addressed outside of science (in a government document, for example)" (p. 211).

Our analyses revealed the challenges in measuring policy impact via citation patterns. In fact, the closer a document is related to actual decision-making the fewer citations it may contain. For example, scientific assessments of the literature contain large numbers of citations, but they are not directly used in policy-making. Instead they are further built upon and "translated" in policy briefs, policy reports, briefing notes or ministerial expertise. The final political decision – usually a legal text – usually does not contain any citations. As we move towards real decisions it therefore gets increasingly challenging to measure impact in this way. Future work may therefore be organized around a simple model of policy impact considering a chain of different document types. Scientific assessment reports, systematic reviews or meta-analyses – as recommended by Bornmann and Marx (2014) – may be the starting point as rigorous syntheses of the available summaries. Next might be science communication documents such as policy briefs, policy reports or plain language summaries. Government reports might be compiled to directly inform particular decisions and, finally, legislative documents cover the policies themselves. In this context, Isett and Hicks (2019) speak about knowledge intermediaries in document chains. Future research could attempt measuring the impact on policy along such a document chain. As citations would be expected to fade away as you move down the chain, it will become increasingly relevant to use text mining methods to measure impact.

Thus, we think that primary research should not be used in policy documents. Unless those documents are "scientific assessments themselves" like those by the IPCC (but even for those the argument holds, see Minx, Callaghan, Lamb, Garard, & Edenhofer, 2017), they should mainly be built from secondary research that uses formal research synthesis methods to understand what empirical findings are robust and which ones are not. Single primary studies are not fit for policy advice as they depend too immediately on the particular research design, datasets used and method of analysis. From a science-policy perspective it is therefore of immediate interest to understand how primary and secondary research evidence are used in local, national and international climate policies – another strand of future research.

## Acknowledgements

The bibliometric data used in this paper are from an in-house database developed and maintained by SciTech Strategies, Inc. derived from Scopus, prepared by Elsevier BV (Amsterdam, The Netherlands). The policy document data were shared with us by Overton on December 04, 2020.